\documentclass[conference]{IEEEtran}
\IEEEoverridecommandlockouts
\usepackage{flushend,cite}
\usepackage{amsmath,amssymb,amsfonts}
\usepackage{algorithmic}
\usepackage{graphicx}
\usepackage{textcomp}
\usepackage{xcolor}
\usepackage{booktabs}
\usepackage{subfigure}

\usepackage{placeins}
\usepackage{float}
\restylefloat{figure}

\usepackage{sidecap}

\usepackage[hyphens]{url}
\usepackage{hyperref}
\usepackage{booktabs}
\usepackage{tikz}
\usetikzlibrary{positioning}
\usetikzlibrary{shapes.geometric}

\usepackage{balance}

\def\BibTeX{{\rm B\kern-.05em{\sc i\kern-.025em b}\kern-.08em
    T\kern-.1667em\lower.7ex\hbox{E}\kern-.125emX}}
\begin{document}

\title{Estimation of Ground NO2 Measurements\\ from Sentinel-5P Tropospheric Data\\ through Categorical Boosting
}
\makeatletter
\newcommand{\newlineauthors}{%
  \end{@IEEEauthorhalign}\hfill\mbox{}\par
  \mbox{}\hfill\begin{@IEEEauthorhalign}
}
\makeatother

\author{\IEEEauthorblockN{1\textsuperscript{st} Francesco Mauro}
\IEEEauthorblockA{\textit{Engineering Department} \\
\textit{University of Sannio}\\
Benevento, Italy  \\
f.mauro$@$studenti.unisannio.it}
\and
\IEEEauthorblockN{2\textsuperscript{nd} Luigi Russo}
\IEEEauthorblockA{\textit{Engineering Department} \\
\textit{University of Sannio}\\
Benevento, Italy \\
l.russo15@studenti.unisannio.it}
\and
\IEEEauthorblockN{3\textsuperscript{rd} Fjoralba Janku}
\IEEEauthorblockA{\textit{Engineering Department} \\
\textit{University of Sannio}\\
Benevento, Italy \\
f.sota@studenti.unisannio.it}
\and
\newlineauthors

\IEEEauthorblockN{4\textsuperscript{th} Alessandro Sebastianelli}
\IEEEauthorblockA{\textit{$\phi$-lab} \\
\textit{European Space Agency}\\
Frascati, Italy  \\
Alessandro.Sebastianelli@esa.int}
\and
\IEEEauthorblockN{5\textsuperscript{th} Silvia Liberata Ullo}
\IEEEauthorblockA{\textit{Engineering Department} \\
\textit{University of Sannio}\\
Benevento, Italy \\
ullo@unisannio.it}
}

\maketitle

\begin{abstract}
This study aims to analyse the Nitrogen Dioxide ($\mathbf{NO_2}$) pollution in the Emilia Romagna Region (Northern Italy) during 2019, with the help of satellite retrievals from the {\nobreak Sentinel\nobreak-5P} mission of the European Copernicus Programme 
and ground-based measurements, obtained from the ARPA site (Regional Agency for the Protection of the Environment). 
The final goal is the estimation of ground $\mathbf{NO_2}$  measurements when only satellite data are available. 
For this task, we used a Machine Learning (ML) model, Categorical Boosting, which  was demonstrated to work quite well and allowed us to achieve a Root-Mean-Square Error ($RMSE$) of 0.0242 over the  43 stations utilised to get the Ground Truth values. 
This procedure represents the starting point to understand which other actions must be taken to improve the final performance of the model and extend its validity.
\end{abstract}


\section{Introduction}

Atmospheric pollution has been largely considered by the scientific community as a primary threat to human health and ecosystems, above all for its impact on climate change. Therefore, its  containment and reduction are gaining interest and commitment from institutions and researchers, although the solutions are not immediate.  It becomes of primary importance to identify the distribution of air pollutants and evaluate their concentration levels in order to activate the right countermeasures. 
One of the most dangerous pollutants in the atmosphere is the Nitrogen Dioxide ($NO_2$) \cite{logan1983nitrogen}, since it 
causes respiratory and cardiovascular diseases, weakening  the immune function of the lungs, such as  asthma and many others illnesses \cite{latza2009effects}.
In the past, the Ozone Monitoring Instrument (OMI) \cite{amt-10-3133-2017}
and Global Ozone Monitoring Experiment (GOME) \cite{RICHTER20021673} satellite sensors were used  to measure $NO_2$ pollution, but since October 2017 the Sentinel-5P mission of the European Space Agency (ESA) Copernicus program has transformed the air quality monitoring and data collection
mainly due to the high resolution of the on-board instrument, enabling investigations  at a local and large scale \cite{ijerph18020544, Sentinel5PoverEurope}.\\
\begin{figure*}[!ht]
    \centering
    \subfigure[]{\includegraphics[width=0.45\textwidth]{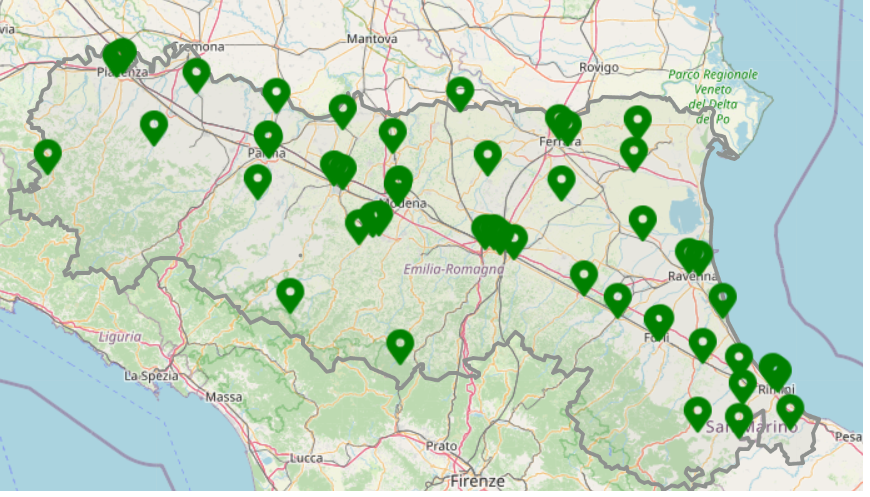}} 
    \subfigure[]{\includegraphics[width=0.45\textwidth]{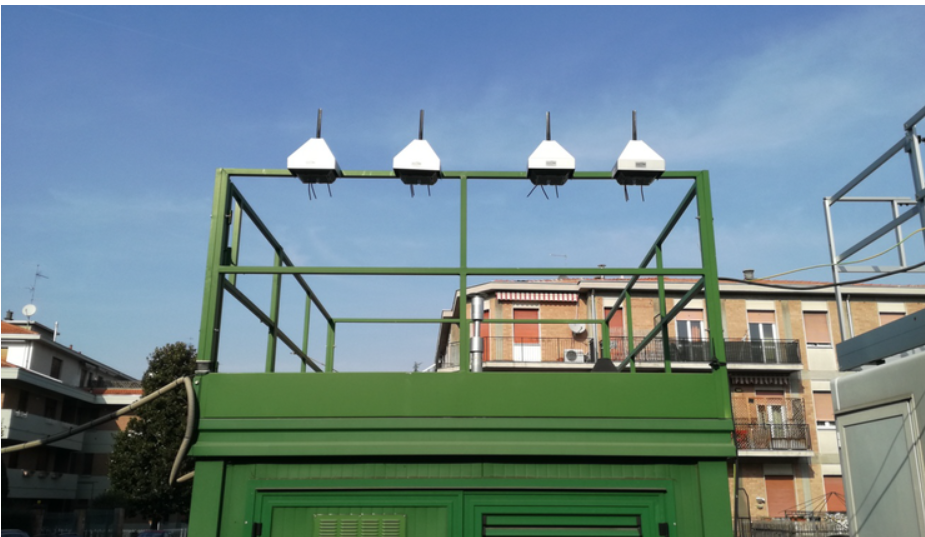}} 
    \caption{(a) Map of fixed stationsin Emilia-Romagna region. (b) Sensors installed in Arpae air quality monitoring unit \\(source: \cite{anagrafeStazioni}, \cite{storicoStazioni}, ARPAE)}
    \label{fig:foobar}
\end{figure*}
In addition, the advance of Machine Learning (ML) models  and their application to Remote Sensing (RS) \cite{rosso2021artificial}
have enabled more accurate estimation of different pollutants compared to traditional statistical algorithms \cite{CHEN2019556}, \cite{XU202157}, and specifically, mod\nobreak els such as Neural networks \cite{CHEN2019104934}, Bayesian maximum entropy \cite{jiang2018space}, random forest \cite{zhan2018satellite} have outcome the traditional ones. \\
However, many of these algorithms are general and do not take into account the spatio-temporal variations of $NO_2$ and other influencing factors, especially at ground level.
An integration of satellite data and ground measurements have been carried out in works such as \cite{li2020remote}, \cite{wang2018deep} and \cite{ialongo2020comparison}, just to mention a few. In \cite{li2020remote} the authors use a geographically and temporally weighted generalised regression neural network (GTW-GRNN) to estimate ground $NO_2$ concentrations by
integrating ground $NO_2$ station measurements, $NO_2$ remote sensing products, meteorologic data, and other auxiliary data. The second paper  \cite{wang2018deep} deals with spatial correlations between the stations, included in a deep learning architecture, and the temporal parameters are modelled with Long Short-Term Memory (LSTM) layers. 
In \cite{ialongo2020comparison} the authors  describe a comparison of  $NO_2$ measurements made by the TROPOMI/Sentinel-5P satellite and those made by air monitoring stations in Helsinki, Finland.

Moreover, in recent years, $NO_2$ concentrations variations have gained a lot of attention and different studies have been conducted using TROPOMI instrument regarding the reduction of air pollutants during the COVID-19 pandemic period \cite{sebastianelli2021airsense}. The effect of lockdown on $NO_2$ in London, Milan and Paris was studied in \cite{COLLIVIGNARELLI2021952}  and the descriptive and quantitative analysis results in a high decrease of $NO_2$ due to low traffic. Similar researches are done on air pollutants such as $PM2.5$, $PM10$, $CO$ and $NO_2$ in various regions of Europe and United States of America (USA). Satellite and ground-based data indicate that limited human interaction in the natural environment has brought on COVID pandemic a positive impact\cite{SANNIGRAHI2021110927, rs12101613}.

In our study, the potential of a specific ML approach is analysed, to verify its ability to reconstruct  the correlation between the variation of the average concentrations of $NO_2$ at the tropospheric level and the punctual one, respectively detected by Sentinel-5P and the ground stations on the national territory of Italy. The idea is to find a transfer-model able to furnish in output ground-level values of the pollutants when only satellite data are available. In this work, the attention has been restricted to the Emilia Romagna region, in Italy, but future works will focus on its extendability.

\section{Data sources}
Satellite data from the Copernicus Sentinel-5P mission 
and ground-based measurements, obtained from 43  ARPA stations in Emila-Romagna region, are utilized  \cite{anagrafeStazioni}, \cite{storicoStazioni}.

\subsection{Sentinel-5P}
The European Earth Surface Observation Program is based on a series of six satellites, called Sentinels, designed by ESA. Among the priority applications declared in the program, there are climate change management, ocean monitoring, atmospheric monitoring, terrestrial monitoring, environmental security, and emergency monitoring. In our work,  particular attention is paid to Sentinel–5P, whose data are used in the experimental part of the study. 
The Sentinel-5P was launched in October 2017 and supports the global monitoring of the atmosphere and air quality using TROPOMI. The TROPOMI is a passive instrument and it consists of four spectrometers measuring radiation in the ultraviolet (270–320 $nm$), visible (310–500 $nm$), near-infrared (675–775 $nm$), and shortwave infrared (2305–2385 $nm$) electromagnetic spectrums. At the beginning of the mission, the spatial resolution of the TROPOMI products was 3.5 $km$ × 7 $km$, while currently (December 2022) it is 3.5 $km$ × 5.5 $km$. TROPOMI has two kinds of tropospheric $NO_2$ products. One is a near-real-time product, which allows users to obtain the imaging data of an area within 4 hours after the satellite has scanned that area and the second is an offline product that allows satellite data to be downloaded from the official website within a few days \cite{amt-10-3133-2017}. \\
In this study, we use TROPOMI offline version of the tropospheric $NO_2$  column concentration L2-type product based on the inversion of differential absorption spectroscopy released by ESA. 
Google Earth Engine (GEE)
platform \cite{GEE} was used to select and download the data.

\subsection{ARPA stations}
Air quality monitoring is performed in the Emilia Romagna region by the Regional Environmental Protection Agency\\ (
ARPA Emilia Romagna,  ARPAE
\cite{ARPAE}) through its air quality sensors network. The actual network is composed of 47 fixed ground stations (see Figure 1) providing hourly ground-level measures of pollutants concentrations [$\mu g/m^3$], including $NO_2$, $PM10$, $PM2.5$, $O_3$, and $SO_2$. $NO_2$ hourly concentrations for the analysis periods  and the sensors metadata  have been  downloaded as CSV file and 43 of the 47 sensors were considered, according to the availability of complete time-series measurements.

\section{Area of Interest and ML model}
Monitoring of the $NO_2$ pollution in the Emilia Romagna Region (Northern Italy) during the period January-July 2019 has been carried out.   
In  Figure 2,  we have shown the map of a $NO_2$ tropospheric vertical column over Italy, with higher values highlighted in red and lower in light blue, where the 
Panoply software \cite{Panoply}
was used to create the related image. Panoply is a very useful free tool to open netCDF4 files that contain Sentinel-5P data.
\begin{figure}[!ht]
	\centering
    \includegraphics[width= 0.47\textwidth]{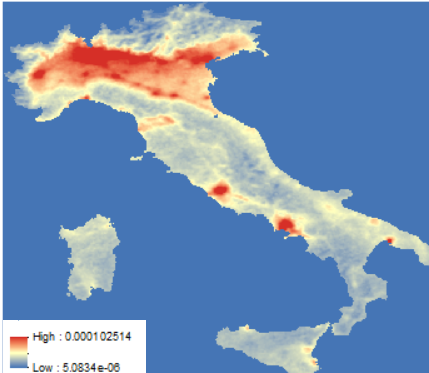}
	\caption{ Sentinel - 5P Nitrogen dioxide in Italy - May to August 2019 ($mol/m^2$)}
	\label{fig:nitrogen dioxide}
\end{figure}

As underlined before, the final goal of our research is the estimation of ground $NO_2$  measurements when only satellite data are available. To this end, an ML model, the Categorical Boosting (CatBoost) \cite{dorogush2018catboost}, has been used to find the correlation between the tropospheric $NO_2$  and the $NO_2$  on the ground. In the ML field,  Boosting can be defined as a set of algorithms whose main function is to convert weak learners into strong learners, where a learner refers to the part of the code that extracts the ML model from the data. This technique applied to algorithms such as Random Forest, Decision Tree or Logistic Regression can help them to overcome simpler algorithms. Indeed, they can considerably improve the prediction accuracy of the model. In the following a brief explanation of CatBoost functioning is given. 
In CatBoost the obtained enhancement combines weak learners to form a strong learner. Unlike a weak learner, a strong learner is a classifier associated with the correct categories.
Suppose we create a Random Forest model that gives 75 \%  accuracy on the validation data set, and later we decide to try another model on the same data set. Assuming that we test as other models, the linear regression and the K-nearest neighbors (kNN) model on the same set of validation data, and we verify that the model is able to offer an accuracy of respectively 69\% and 92\%, it is clear that all three models work in completely different ways and give completely different results on the same data set.    We can highlight that CatBoost significantly improves the accuracy of any model’s prediction, allowing to reach a comparable value, because it converts weak learners into strong learners, where learners mean the part of code that extracts the ML model from data. To identify weak learners, ML algorithms are used with a different distribution for each iteration and  a new weak prediction rule is generated for each algorithm. After several iterations, the boosting algorithm combines all weak learners to form a single prediction rule \cite{CatBoosting}. 
Thus, the CatBoost is an algorithm 
based on binary decision trees \cite{breiman2017classification}\cite{naseer2022machine}, owning the characteristics to divide the original features space $R^m$ into leaves, with a constant value in each region, according to the values of some splitting attributes. 

We chose it because it is a powerful ML model, as demonstrated by many researches, especially for heterogeneous features, noisy data, complex dependencies and in addition, it treats categorical feature transformation better than the other ML models\cite{dorogush2018catboost} \cite{HUANG20191029}\cite{JABEUR2021120658}.

The model was  created  through  the  CatBoostRegressor class, and the relative parameters were defined, like for instance the number of iterations, learning rate, loss function, total number of epochs, the $max\_depth$ parameter (whose limit value is fixed to 16).  
Then the CatBoost was trained and tested on the related dataset. In Figure \ref{fig:catboost}, the scheme of the CatBoost used in our case is shown.
\begin{figure}[H]
	\centering
\includegraphics[width= 0.52\textwidth]{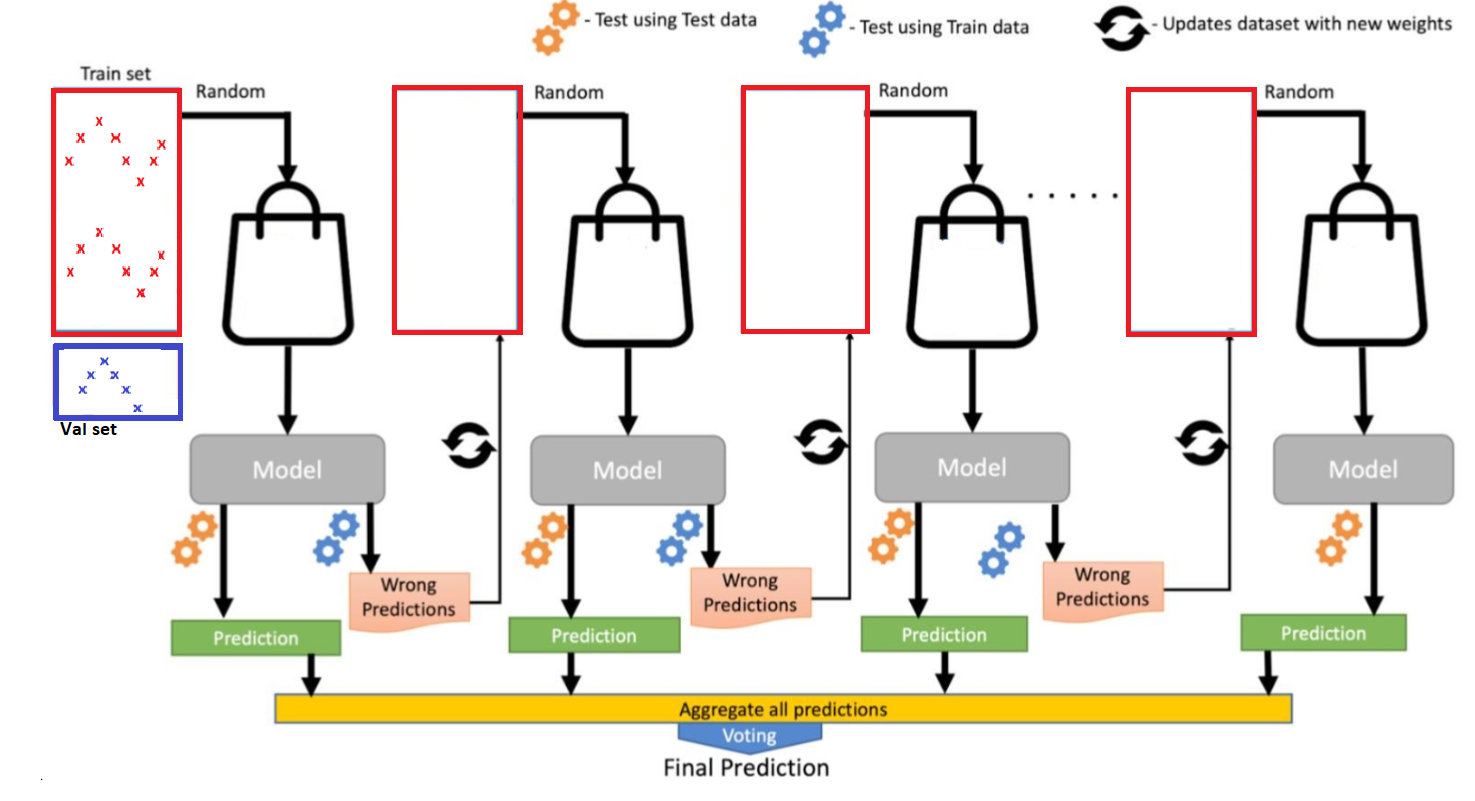}
	\caption{The scheme of CatBoost used in our case.}
	\label{fig:catboost}
	\end{figure}
\section{Dataset Creation and Methodology}
\subsection{Dataset}
As anticipated in the previous section, Sentinel-5P data and ground-based measurements were  utilised to train the CatBoost and to find the correlation among the different $NO_2$ measurements. We first proceed with the construction of the dataset. Through the corresponding station code in the ARPAE registry, the historical values referred to the years 2010-2020 were analysed, which contain the  values of $NO_2$ pollutant (unit of measurement µg/$m^3$), measured hour by hour. 
Only the values in the same time interval of analysis considered for the GEE Sentinel-5P data (1 January 2019 - 30 July 2019) were then considered. By way of example in Table 1 the ARPAE values of a specific station are plotted, collected over some hours of January $1^{st}$ 2019. 
\subsection{Methodology}
Data manipulation was carried out to aggregate the Sentinel-5P data, which have a daily temporal resolution, to the measurements taken on the ground, which instead vary hour by hour. On a practical level, this was done by averaging all the values of the soil pollutant measurements with hourly time resolutions, and then by comparing this result with the respective Sentinel-5P data. 
\begin{table}
    \centering
    \maketitle
Table 1: $N0_2$ values from one of the ARPAE stations\\ over the $1{st}$ of January 2019\\
\vspace{0.6cm}
    \begin{tabular}{|l|l|l|l|l|l|}
    \hline
        CO\_STA &ID & DA\_INIZ & DA\_FIN & VAL & UM \\ \hline
        10000074 & 8 & 01/01/2019 02 & 01/01/2019 03 & 7 & ug/m3 \\ \hline
        10000074 & 8 & 01/01/2019 03 & 01/01/2019 04 & 5 & ug/m3 \\ \hline
        10000074 & 8 & 01/01/2019 04 & 01/01/2019 05 & 4 & ug/m3 \\ \hline
        10000074 & 8 & 01/01/2019 05 & 01/01/2019 06 & 5 & ug/m3 \\ \hline
        10000074 & 8 & 01/01/2019 06 & 01/01/2019 07 & 5 & ug/m3 \\ \hline
        10000074 & 8 & 01/01/2019 07 & 01/01/2019 08 & 4 & ug/m3 \\ \hline
        10000074 & 8 & 01/01/2019 08 & 01/01/2019 09 & 4 & ug/m3 \\ \hline
        10000074 & 8 & 01/01/2019 09 & 01/01/2019 10 & 4 & ug/m3 \\ \hline
        10000074 & 8 & 01/01/2019 10 & 01/01/2019 11 & 4 & ug/m3 \\ \hline
        10000074 & 8 & 01/01/2019 11 & 01/01/2019 12 & 5 & ug/m3 \\ \hline
        10000074 & 8 & 01/01/2019 12 & 01/01/2019 13 & 5 & ug/m3 \\ \hline
        10000074 & 8 & 01/01/2019 13 & 01/01/2019 14 & 7 & ug/m3 \\ \hline
        10000074 & 8 & 01/01/2019 14 & 01/01/2019 15 & 7 & ug/m3 \\ \hline
        10000074 & 8 & 01/01/2019 15 & 01/01/2019 16 & 9 & ug/m3 \\ \hline
        10000074 & 8 & 01/01/2019 16 & 01/01/2019 17 & 10 & ug/m3 \\ \hline
        10000074 & 8 & 01/01/2019 17 & 01/01/2019 18 & 8 & ug/m3 \\ \hline
        10000074 & 8 & 01/01/2019 18 & 01/01/2019 19 & 6 & ug/m3 \\ \hline
        10000074 & 8 & 01/01/2019 19 & 01/01/2019 20 & 6 & ug/m3 \\ \hline
        10000074 & 8 & 01/01/2019 20 & 01/01/2019 21 & 6 & ug/m3 \\ \hline
        10000074 & 8 & 01/01/2019 21 & 01/01/2019 22 & 6 & ug/m3 \\ \hline
        10000074 & 8 & 01/01/2019 22 & 01/01/2019 23 & 6 & ug/m3 \\ \hline
        10000074 & 8 & 01/01/2019 23 & 02/01/2019 00 & 7 & ug/m3 \\ \hline
        10000074 & 8 & 02/01/2019 00 & 02/01/2019 01 & 10 & ug/m3 \\ \hline
        10000074 & 8 & 02/01/2019 01 & 02/01/2019 02 & 9 & ug/m3 \\ \hline
        10000074 & 8 & 02/01/2019 02 & 02/01/2019 03 & 7 & ug/m3 \\ \hline
        10000074 & 8 & 02/01/2019 03 & 02/01/2019 04 & 5 & ug/m3 \\ \hline
        10000074 & 8 & 02/01/2019 13 & 02/01/2019 14 & 13 & ug/m3 \\ \hline
        10000074 & 8 & 02/01/2019 14 & 02/01/2019 15 & 10 & ug/m3 \\ \hline
        10000074 & 8 & 02/01/2019 15 & 02/01/2019 16 & 9 & ug/m3 \\ \hline
        10000074 & 8 & 02/01/2019 16 & 02/01/2019 17 & 8 & ug/m3 \\ \hline
        10000074 & 8 & 02/01/2019 17 & 02/01/2019 18 & 7 & ug/m3 \\ \hline
        10000074 & 8 & 02/01/2019 18 & 02/01/2019 19 & 9 & ug/m3 \\ \hline
        
         \end{tabular}  
\end{table}
\vspace{4 cm}
For the Sentinel-5P data, two types of inputs have been  created.
The first one contains  the $NO_2$ values expressed in the original unit of measurement $mol/m^2$ and the second one expressed in $\mu g/ m^3$ to match the data from the ground stations. 

The conversion was done using a climatological model \cite{NO2convertion} which consists of dividing the tropospheric  value of $NO_{2}$ by the height of the air column (which can vary from a minimum of 8 $km$ to a maximum of 20 $km$ then multiplying it by a factor of one thousand and by the molar mass of nitrogen dioxide itself (equal to a value of 46,055 $g/mol$). However, the limit of this approximation must be taken into account, since the column should  contain the gas homogeneously (it is, therefore, an auxiliary information to be given to the model).
The two types of input were also pre-processed through the use of the \textit{Savitzky-Golay} filter chosen because
performs an appropriate data smoothing by removing the spurious components and maintaining however the main features \cite{schafer2011savitzky}. 
The same filtering was also applied to ground data, which represent the ground truth (GT), since experiments have shown that the model works better when this filter is employed. 
Then we proceed with a shuffle operation of the data which consists in mixing the combination of Sentinel-5P  data as described above and the GT data, by maintaining their correspondence. This allows the model to free itself from unwanted patterns. Lastly, a  splitting is carried on, and  80\% of the total data are assigned to the training set and 20\% to the test set. 
\vspace{0.7cm}
\section{First results}
 Some trials were carried on  to adjust the CatBoost parameters. We started  by modifying the value of the $max\_depth$, and by setting it to three possible values: 3, 6, and 12.
 After verifying that the best performance is obtained with a $max\_depth = 12$, this value was fixed, 
and some other simulations were carried out by acting on the height (h) of the gas column, containing the tropospheric $NO_2$ values, by using a minimum value of 8 $km$, an intermediate value of 13 $km$, and a maximum value of 20 $km$. Finally, after training the model, its performance was evaluated by means of a comparison, for every single station, between the GT and the Model Prediction (MP). 

\subsection{Quantitative analyses}
\begin{figure*}[ht!]
	\centering
\includegraphics[width= 0.80\textwidth]{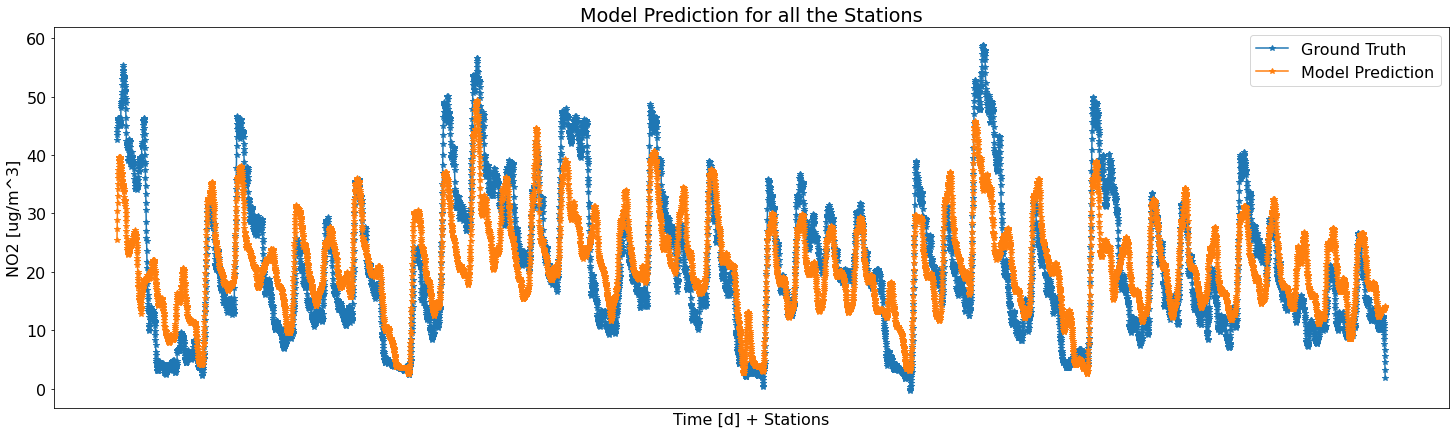}
	\caption{Model Prediction for all the stations when $max\_depth = 12$, and h = 13 km.}
	\label{fig:maxDepthoptimum}
	\end{figure*}
The statistical metric used in this study to measure the prediction performance of the developed model is the Mean Square Error ($RMSE$). 
$RMSE$ as defined by equation (1) is a parameter that measures the difference between the predicted value of the model and the real value. It is very sensitive to extremely large or small errors in a set of data, so it can reflect well the accuracy of the real value.

\begin{equation} \label{eq3}
RMSE= \sqrt{ \frac{1}{m} \sum_{i=1}^m (y'_i - y_i)^2}
\end{equation}

Table 2 shows the mean values (for all the stations) obtained for the specific metric when $max\_depths$ varies. 
\begin{center}
\centering
\maketitle
Table 2: Averaged $RMSE$ on the validation set\\ for different $max\_depths$ \\
\vspace{0.6cm}
\begin{tabular}{ccc} 
\toprule
 Max depth = 3 & Max depth = 6 & Max depth = 12 \\
 \midrule
 0.1039 & 0.1037 & 0.0242 \\  
 \bottomrule
\end{tabular}
   
\end{center}

Table 3 shows the RMSE when h varies and it is equal to 8 $km$, 13 $km$ and 20 $km$. 

\begin{center}
\centering
\maketitle
Table 3: Averaged $RMSE$ on validation set for different values of the gas column height ($h$).

\vspace{0.5cm}
\begin{tabular}{ccc}
    \toprule
  $h$ = 8 km & $h$ = 13 km  & $h$ = 20 km \\
 \midrule
 0.1080 & 0.0242 & 0.1057 \\  
 \bottomrule
  \end{tabular}
\end{center}

\vspace{1cm}
From the simulations carried out, it emerges that the config\nobreak uration that guarantees the best performance is the one with $max\_depth = 12$ and $h$ = 13 km. In fact, in this case we have the lowest average value of $RMSE$ calculated among all the 43 stations. 
For this combination of $max\_depth$ and $h$, the model prediction is shown in Figure \ref{fig:maxDepthoptimum} (orange graph), and compared with the GT (blue graph). The two trends appear to be very similar and the values are consistent with each other.

However, further work is necessary to verify its trans\nobreak ferability to other areas of interest and to improve its performance.

\section{Conclusions}
This study aimed to analyse the $NO_2$ pollution in the Emilia Romagna Region (Northern Italy),  with the help of satellite retrievals from the Sentinel-5P mission of the European Copernicus Programme over the period 1 January 2019 - 30 July 2019,  and through ground-based measurements, obtained from the ARPA site of the same region.

It was found that the CatBoost  model  gives  the best value of Root-Mean-Square Error ($RMSE$ = 0.0242) for $max\_depth = 12$ and $h$ = 13 km when the  43 stations are considered.  This aspect is confirmed by the fact that the trends of ground truth and Model Prediction appear to be very similar and while  slightly distant their values are consistent with each other. 
The conclusions drawn from this study will be  used in future work to improve the model and to verify its accuracy for other areas of Italy and the world and analyse other strategies to further improve its performance. 
\bibliographystyle{IEEEtran}
\bibliography{ref}
\end{document}